\documentclass[aip,jcp,reprint,numerical,english,floatfix]{revtex4-1}
\usepackage{xcolor}
\usepackage{pdfcolmk}
\usepackage{amsmath}
\usepackage{amssymb}
\usepackage{graphicx}
\usepackage{esint}
\usepackage{ulem}

\usepackage{array}
\usepackage{bm}
\usepackage{amsmath}
\usepackage{graphicx}
\usepackage{babel}
\usepackage{babel}

\begin{document}
\global\long\def\deriv#1#2{\frac{\mathrm{d}#1}{\mathrm{d}#2}}
\global\long\def\pderiv#1#2{\frac{\partial#1}{\partial#2}}
\global\long\def\e{\mathrm{e}}
\global\long\def\d{\mathrm{d}}
\global\long\def\vec#1{\bm{#1}}
\global\long\def\icm{\mathrm{\mathrm{cm}^{-1}}}
\global\long\def\imath{\mathrm{i}}
\global\long\def\ptderąiv{\frac{\partial}{\partial t}}
\renewcommand{\figurename}{Fig.}

\title{Molecular vibrations-induced quantum beats in two-dimensional electronic
spectroscopy}

\author{Vytautas Butkus}

\affiliation{Department of Theoretical Physics, Faculty of Physics, Vilnius University,
Sauletekio 9-III, 10222 Vilnius, Lithuania}

\affiliation{Center for Physical Sciences and Technology, Gostauto 9, 01108
Vilnius, Lithuania}

\author{Leonas Valkunas}

\affiliation{Department of Theoretical Physics, Faculty of Physics, Vilnius University,
Sauletekio 9-III, 10222 Vilnius, Lithuania}

\affiliation{Center for Physical Sciences and Technology, Gostauto 9, 01108
Vilnius, Lithuania}

\author{Darius Abramavicius}

\email{darius.abramavicius@ff.vu.lt}

\affiliation{Department of Theoretical Physics, Faculty of Physics, Vilnius University,
Sauletekio 9-III, 10222 Vilnius, Lithuania}

\affiliation{State Key Laboratory of Supramolecular Complexes, Jilin University,
2699 Qianjin Street, Changchun 130012, PR China}
\begin{abstract}
Quantum beats in nonlinear spectroscopy of molecular aggregates are
often attributed to electronic phenomena of excitonic systems, while
nuclear degrees of freedom are commonly included into models as overdamped
oscillations of bath constituents responsible for dephasing. However,
molecular systems are coupled to various high-frequency molecular
vibrations, which can cause the spectral beats hardly distinguishable
from those created by purely electronic coherences. Models containing
\emph{damped, undamped} and \emph{overdamped} vibrational modes coupled
to an electronic molecular transition are discussed in this paper
in context of linear absorption and two-dimensional electronic spectroscopy.
Analysis of different types of bath models demonstrates how do vibrations
map onto two-dimensional spectra and how the damping strength of the
coherent vibrational modes can be resolved from spectroscopic signals. 
\end{abstract}
\maketitle

\section{Introduction}

Various ultrafast spectroscopy techniques are capable of probing molecular
dynamics on timescales of nuclear motion and spectroscopic signals
usually reveal effects caused by the coherent exciton dynamics, energy
transfer, vibrational motions, etc. perceptible even for simple systems
\cite{Kim2009,TianKeustersSuzakiEtAl2003,engel-nat2007,mancal-sperling-jcp2010,Abramavicius2010}.
For example, coherent quantum beats obtained at broadband excitation
conditions are displayed in spectra of small molecular complexes \cite{Okamoto1991,cheng-fleming-JPCA08,Perakis2011},
thus, reflecting an interplay between the absorbed energy transfer
and relaxation.

The two-dimensional (2D) electronic spectroscopy (ES) displays highly
resolved decoherence dynamics in molecular aggregates induced by the
system-bath interaction. Problem of assignment of the detected slowly-decaying
spectral beats to either electronic coherences \cite{engel-nat2007,Collini-Scholes-SCI2009,caruso-plenio-JCP09,PanitchayangkoonEngel2010,read-pnas2007,ColliniScholes2010}
or molecular vibrations \cite{Bixon1997,Vos1991} is under intense
discussions up to now. However, it is still disputed whether such
detachment is feasible at all.

Molecules and their environment are characterized by a wide variety
of high-frequency vibrational modes, some of them strongly-coupled
to the electronic excitations. Unambiguous experimental evidences
that the high-frequency vibrational wavepacket motion causes the spectral
beats in 2D spectra were reported only recently \cite{nemeth-sperling-JCP2010,Christensson2011,Turner2011}.
However, it is not yet clearly established how these vibrational resonances
map onto 2D spectra and how the excited vibrational wavepacket
evolution affects the spectral dynamics. 

Systems where the high-frequency vibrations are mixed with electronic
transitions are currently of a special interest. Great efforts are
made to effectively and universally identify the underlying mechanisms
\cite{Turner2011,Christensson2011,Butkus2012,Caycedo-Soler2012}.
Analysis of \emph{rephasing} and \emph{nonrephasing} signals of the
2D ES separately could help solving this issue in some cases, but
usually the strong spectral congestion leaves ambiguity of the result
due to the overlap of many oscillating features. In case when the
limited-bandwidth laser pulses cannot cover the whole absorption spectrum
even if the vibrational features are well-resolved, the intramolecular
vibrations can still be evaluated by analyzing the phase relationships
of beating patterns in the rephasing and nonrephasing signals, peak
ellipticity or rotation of the nodal line of the imaginary part of
the total 2D spectrum \cite{Mancal_JPC2012,Butkus2012,mancal-sperling-jcp2010}.
\textcolor{black}{However, vibrational coherences can be created in
the electronic ground state manifold resulting in beats of both diagonal
and off-diagonal peaks, thus making the whole time-resolved dynamics
even more entangled \cite{Mancal_JPC2012,Butkus2012}. }

Non-decaying vibrations-related quantum beats on top of the vertical
electronic transitions are obtained in simulated time-resolved spectra
when molecular vibrations are included as coherent harmonic modes
of the bath coupled to the system \cite{mancal-sperling-jcp2010,Christensson_JPCB2012,Womick2011}.\textcolor{black}{{}
}This problem sometimes is formulated in pure quantum mechanical terminology
by using so-called polaron transformation \cite{Holstein1959,Cheng2008c}.\textcolor{black}{{}
Such treatment is exact when non-interacting molecules are considered;
however, in molecular aggregates vibrations are coupled to the electronic
degrees of freedom in a non-trivial way. More sophisticated theoretical
methods, for example, hierarchical equations of motion or time-dependent
density matrix renormalization group technique, might resolve the
dynamics of such mixed vibronic systems \cite{Zhu2011,Hu2011,Prior2010}.}
As the vibrational coherence decay rate is usually smaller compared
to those of the electronic coherences, this might also trigger the
reconsideration of the nature of long-lived coherences observed in
many systems.

In this paper, we theoretically analyze the influence of high-frequency
\emph{damped} vibrational motion of molecules on the spectroscopic
observables and make comparison with the \emph{undamped} description.
In the microscopic model such treatment translates into a more sophisticated
expression of bath spectral density function. General bath damping
mechanisms are also the problem under consideration; the model of
a \emph{quantum overdamped }bath is introduced and its influence upon
the lineshape formation in the linear absorption and 2D spectrum is
discussed. Strongly-overlapping spectra of simple electronic and vibrational
systems are examined in the context of various bath models.

\section{Generic molecular complexes}

Electric fields interacting with molecular systems can affect various
system degrees of freedom ranging from electronic high-energy states
to low-frequency molecular vibrations. It is natural to simplify the
description by taking into consideration only the most relevant types
of dynamics. One of the main simplification is achieved by neglecting the 
non-resonant transitions with respect to the optical field under consideration:
in the case of the resonant excitation with the optical laser field
a molecule is described as a system consisting of two electronic energy
levels, while a supramolecular complex -- as an aggregate of coupled
two-level electronic systems. Another standard assumption when describing
electronic quantum levels of a molecule is the separation of electronic
and nuclear degrees of freedom (the Born-Oppenheimer approximation).
It is assumed that the electronic wavefunction rigorously follows
the nuclear configuration and the energies of electronic levels become
parametric functions of nuclear degrees of freedom. The equilibrium
nuclear configuration which is usually different for each electronic
quantum state is defined as the energy minimum with respect to the
nuclei \cite{valkunasbook,MayBook2011}.

\subsection{\label{sub:MONOMER_VIBRATIONS}Absorption of an isolated molecule}

\begin{figure}
\noindent \begin{centering}
\includegraphics{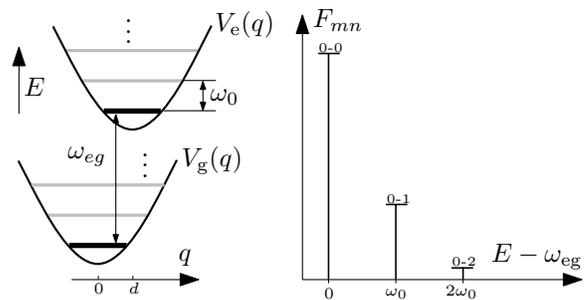} 
\par\end{centering}

\caption{\label{fig:Energy-level-structure}Model system of displaced harmonic
oscillators and Franck-Condon overlap factors for the corresponding
$m-n$ transitions.}
\end{figure}

Using the assumptions made above, electronic excitations in molecules
are approximated as the Franck-Condon transitions. It is considered
that the (slow) nuclear degrees of freedom remain frozen during the
electronic transition and the fixed molecular nuclear configuration
adjusted to the ground electronic state emerges in non-equilibrium
conditions with respect to the new electronic state. As the electronic
potential energy surface is assumed to be parabolic depending on the
nuclear displacement in the vicinity of the equilibrium, a displaced
(harmonic) oscillator model can be used in the description of the
electronic transition (Fig.~\ref{fig:Energy-level-structure}). In
one dimension the electronic potential of the ground state is $V_{\mathrm{g}}(q)=\hbar\omega_{0}q^{2}/2$
and the displaced electronic excited state is described by the potential
$V_{\mathrm{e}}(q)=\omega_{\mathrm{eg}}+\hbar\omega_{0}(q-d)^{2}/2$.
Here $\omega_{0}$ is the vibrational frequency, $\omega_{\mathrm{eg}}$
is the energy gap between the minima of two potentials and $d$ is
a dimensionless displacement parameter determining the strength of
the electron coupling with molecular vibrations. The Huang-Rhys factor
$s=\frac{1}{2}d^{2}$ is widely used to qualify the coupling strength.

Quantum mechanical problem of the isolated harmonic potentials has
exact solutions. They constitute an infinite set of wavefunctions
$\psi_{m}$ with quantum number $m=0\ldots\infty$ and corresponding
energies $E_{m}=\hbar\omega_{0}(m+1/2)$ with respect to the bottom
of the corresponding potential surface. Transitions between the sub-states
of the electronic ground state and the ones of the electronic excited
state determine the vibronic progression in the absorption spectrum (Fig.~\ref{fig:Energy-level-structure}).
The intensity of each vibronic peak is scaled by the overlap of vibrational
wavefunctions -- the Franck-Condon overlap -- in the ground and excited
state potentials. Following these assumptions the absorption spectrum
of a single isolated Franck-Condon molecule can be given by 
\begin{eqnarray}
 &  & \kappa_{\mathrm{abs}}^{\mathrm{FC}}(\omega)\propto\omega\sum_{m,n=0}^{\infty}\mathrm{e}^{-\frac{m\omega_{0}}{k_{\mathrm{B}}T}}|F_{mn}|^{2}\label{eq:abs-monomer-fc}\\
 &  & \qquad\times\mathrm{Re}\int_{0}^{\infty}\mathrm{d}t\e{}^{\mathrm{i}(\omega-\omega_{eg})t-\mathrm{i}\omega_{0}(n-m)t-\gamma t},\nonumber 
\end{eqnarray}
 where the line-broadening parameter $\gamma$ is introduced phenomenologically.
$F_{mn}$ is the Franck-Condon wavefunction overlap integral for the
$m$ to $n$ transition 
\begin{equation}
F_{mn}=\left\langle m\right|\exp\left(-\frac{d}{\sqrt{2}}\left(\hat{a}^{\dagger}-\hat{a}\right)\right)\left|n\right\rangle 
\end{equation}
 given in terms of bosonic (vibrational) creation $\hat{a}^{\dagger}$
and annihilation $\hat{a}$ operators \cite{valkunasbook,MayBook2011}.

\subsection{Molecule coupled to the phonon bath}

In order to capture both the vibrational-type set of states of the
energy spectrum and the spectral broadening in an unified model a
more general description of a molecule is needed. The model discussed
above considered the composite electronic+vibrational system as a
closed system. Consequentially, the line-broadening parameter $\gamma$
was included phenomenologically and without more detailed physical
insight.

The total Hamiltonian of a single molecule coupled to the set of phonon
modes (an open quantum system) is given by 
\begin{eqnarray}
\hat{H} & = & 0|g\rangle\langle g|+(\omega_{eg}+\lambda)|e\rangle\langle e|\label{eq:hamiltonianas}\\
 &  & -\sum_{\alpha}\hbar\omega_{0}^{(\alpha)}\sqrt{s_{\alpha}}\left(\hat{a}_{\alpha}^{\dagger}+\hat{a}_{\alpha}\right)|e\rangle\langle e|\nonumber \\
 &  & +\sum_{\alpha}\hbar\omega_{0}^{(\alpha)}\left(\hat{a}_{\alpha}^{\dagger}\hat{a}_{\alpha}+\frac{1}{2}\right).\nonumber 
\end{eqnarray}
 Here the ground state $|g\rangle$ is the reference zero-energy state,
the energy of the excited state $|e\rangle$ is additionally shifted
by the reorganization energy $\lambda=\sum_{\alpha}\hbar\omega_{0}^{(\alpha)}s_{\alpha}$,
while the rest of vibrational degrees of freedom are represented by
a set $\{\alpha\}$ of harmonic oscillators. The system-bath coupling
represented by the third term in Eq.~\eqref{eq:hamiltonianas} introduces
fluctuations into the energy of the electronic excited state due to
low-frequency bath modes at fixed temperature. The most convenient
form to describe such fluctuations is the time correlation function:
\begin{equation}
C(t)=\mathrm{Tr_{B}}\left\{ Q_\mathrm{e}(t)Q_\mathrm{e}(0)\rho_{\mathrm{eq}}\right\} .\label{eq:trace}
\end{equation}
 Here 
\begin{equation}
\rho_{\mathrm{eq}}=\mathcal{Z}^{-1}\exp\left(-\sum_{\alpha}\beta\hbar\omega_{0}^{(\alpha)}\left(\hat{a}_{\alpha}^{\dagger}\hat{a}_{\alpha}+\frac{1}{2}\right)\right)\label{eq:rho}
\end{equation}
 is the normalized thermally equilibrated canonical density operator
of the bath ($\beta^{-1}=k_{\mathrm{B}}T$), and 
\begin{equation}
Q_\mathrm{e}(t)=\sum_{\alpha}\hbar\omega_{0}^{(\alpha)}\sqrt{s_{\alpha}}\left(\hat{a}_{\alpha}^{\dagger}(t)+\hat{a}_{\alpha}(t)\right)\label{eq:Qe}
\end{equation}
 is the fluctuating collective bath coordinate in the Heisenberg representation.
The exact form of the correlation function describes the high-frequency
modes as well. Trace in Eq.~\eqref{eq:trace} amounts to averaging
over the infinite set $\{\alpha\}$ of harmonic oscillators. Inclusion
of eqs. \eqref{eq:rho} and \eqref{eq:Qe} into the trace gives 
\begin{equation}
C(t)=\sum_{\alpha}\hbar^{2}\left(\omega_{0}^{(\alpha)}\right)^{2}s_{\alpha}\left(\coth\frac{\beta\hbar\omega_{0}^{(\alpha)}}{2}\cos\omega_{0}^{(\alpha)}t-\imath\sin\omega_{0}^{(\alpha)}t\right),\label{eq:c_t}
\end{equation}
 which is the well-known form of the two-point correlation function
of generalized bath coordinates \cite{MayBook2011}. The Fourier transform
of the correlation function is a real function 
\begin{equation}
\mathcal{C}(\omega)=\intop_{-\infty}^{\infty}\d t\e^{\imath\omega t}C(t)\equiv\mathcal{C}^{\prime}(\omega)+\mathcal{C}^{\prime\prime}(\omega),\label{eq:fourier}
\end{equation}
 where $\mathcal{C}^{\prime}(\omega)$ and $\mathcal{C}^{\prime\prime}(\omega)$
are even and odd functions of $\omega$. $\mathcal{C}^{\prime\prime}(\omega)$
is the temperature-independent function and is denoted as the spectral
density: 
\begin{equation}
\mathcal{C}^{\prime\prime}(\omega)=-2\int_{0}^{\infty}\sin\omega t\mathrm{Im}C(t)\d t.\label{eq:corr_def}
\end{equation}
 $\mathcal{C}^{\prime}(\omega)$ and $\mathcal{C}^{\prime\prime}(\omega)$
are related by the fluctuation-dissipation theorem 
\begin{equation}
\mathcal{C}^{\prime}(\omega)=\coth\left(\beta\hbar\omega/2\right)\mathcal{C}^{\prime\prime}(\omega).\label{eq:fl-dis}
\end{equation}

To describe the optical properties of the molecule the system polarization
operator $\hat{P}=\left|g\right\rangle \left\langle e\right|+\left|e\right\rangle \left\langle g\right|$
is assumed (the transition dipole strength is taken as unity). It
can be shown that the quantum correlation functions of the polarization
operator representing the spectroscopic observables are exactly given
in terms of the second-order cumulant expansion with respect to the
vibrational modes \cite{MayBook2011}. Therefore, it is exact for
our molecule and we can calculate the shapes of electronic transition
bands and the bath-induced time dependence of the spectrum by using
the formalism of the lineshape functions. The latter directly comes
from the perturbative second-order cumulant expansion of the system
density operator propagation which allows to describe various types
of vibrational baths and include these effects explicitly \cite{mukbook}.
The absorption coefficient is then given by 
\begin{equation}
\kappa_{\mathrm{abs}}^{\mathcal{C}}(\omega)\propto\omega\mathrm{Re}\int_{0}^{\infty}\mathrm{d}t\e{}^{\mathrm{i}(\omega-\omega_{0})t-g(t)}.\label{eq:abs-monomer-g}
\end{equation}
 Here dimensionless lineshape function $g(t)$ is an integral transformation
of the correlation function $C(t)$ of system-bath fluctuations, or
for its Fourier transform (Eq.~\eqref{eq:fourier}), 
\begin{equation}
g(t)\equiv-\frac{1}{2\pi}\intop_{-\infty}^{\infty}\mathrm{d}\omega\frac{\mathcal{C}(\omega)}{\omega^{2}}\left[\exp(-\mathrm{i}\omega t)+\mathrm{i}\omega t-1\right].\label{eq:lineshape}
\end{equation}

Assuming that the system is coupled to a continuous spectrum of bath
frequencies, the correlation function Eq.~\eqref{eq:c_t} can be
calculated for a predefined distribution. We introduce the density
of the Huang-Rhys parameter as a function of vibronic frequency, $\{s_{\alpha}\}\to s(\omega)\mathrm{d}\omega$,
which may have a peak at some dominant normal-mode frequency $\omega_{0}$.
Then, from eqs.~\eqref{eq:c_t} and \eqref{eq:corr_def} the spectral
density is obtained as 
\begin{equation}
\mathcal{C}^{\prime\prime}(\omega)=\pi\hbar^{2}\omega^{2}\left[s(\omega)-s(-\omega)\right].
\end{equation}
For instance, if we consider a vibrational damped mode it will be
represented by a broad peak in the spectral density of the system-bath
coupling. In case of the Gaussian-type coupling 
\begin{equation}
s_{\mathrm{G}}(\omega)=\frac{1}{\sqrt{2\pi}\gamma}\e^{-\frac{(\omega-\omega_{0})^{2}}{2\gamma^{2}}}
\end{equation}
 the spectral density function is 
\begin{equation}
\mathcal{C}_{\mathrm{G}}^{\prime\prime}(\omega)=\lambda\mu\cdot\frac{\sqrt{\pi}}{\sqrt{2}\gamma}\left[\e{}^{-\frac{\left(\omega-\omega_{0}\right)^{2}}{2\gamma^{2}}}-\e^{-\frac{\left(\omega+\omega_{0}\right)^{2}}{2\gamma^{2}}}\right],\label{eq:C_w_total_gauss}
\end{equation}
 while the Lorentzian-type coupling 
\begin{equation}
s_{\mathrm{L}}(\omega)=\frac{1}{\pi}\frac{1}{(\omega-\omega_{0})^{2}+\gamma^{2}}
\end{equation}
 leads to 
\begin{equation}
\mathcal{C}_{\mathrm{L}}^{\prime\prime}(\omega)=\lambda\mu\cdot\frac{4\omega\omega_{0}\gamma}{\left(\omega^{2}-\omega_{0}^{2}-\gamma^{2}\right)^{2}+4\omega^{2}\gamma^{2}}.\label{eq:C_w_total}
\end{equation}
 Here $\lambda$ and $\mu$ are just some scaling constants here inserted
for convenience: the expression of $\mu$ is later chosen in order
for $\lambda$ to be equal to the reorganization energy $\pi^{-1}\int_{0}^{\infty}\frac{\d\omega}{\omega}\mathcal{C}^{\prime\prime}(\omega)$.

By this point, no approximations were applied -- equations \eqref{eq:C_w_total_gauss}
and \eqref{eq:C_w_total} are consistent with the fluctuation-dissipation
theorem. The inverse Fourier transform of them would give the time
correlation functions which in case of Gaussian and Lorentzian couplings
decay as $\mathrm{e}^{-\gamma^{2}t^{2}/2}$ and $\mathrm{e}^{-\gamma t}$,
respectively. The spectral densities include the damping parameter
$\gamma$ and vibrational frequency $\omega_{0}$. By taking various
limits with respect to these parameters, different damping regimes
can be achieved representing different conditions of the bath and
both spectral densities (eqs. \eqref{eq:C_w_total_gauss} and \eqref{eq:C_w_total})
can be used in numerical simulations.

\begin{figure}
\includegraphics{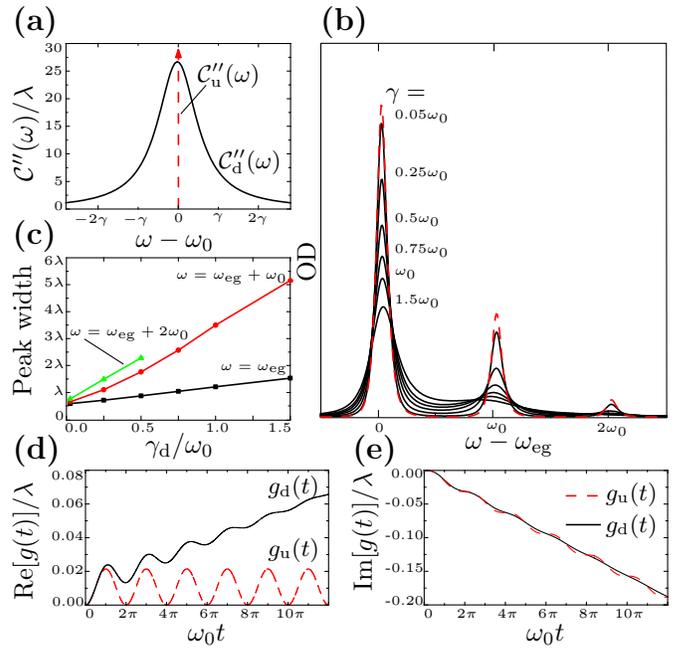}

\caption{\label{fig:Damped_overdamped}(a) Spectral densities of \emph{undamped
}(red dashed line) and \emph{damped} (black solid line) baths. (b)
Absorption spectra of the displaced harmonic oscillator ($s=0.3$)
for different damping strengths $\gamma/\omega_{0}=0.05-1.5$ of the
damped bath (black solid lines) and for the undamped bath (red dashed
line). The widths of three main peaks are displayed as the functions
of damping strengths in (c), where the third peak becomes unresolvable
for $\gamma>0.5\omega_{0}$. The real and imaginary parts of the lineshape
functions corresponding to the different vibrational bath models are
plotted in (d) and (e), respectively.}
\end{figure}

We further analyze different regimes of \emph{undamped}, \emph{damped}
and \emph{overdamped} vibrational motion considering the Lorentzian-type
coupling since it is related to the exponentially-decaying time correlation
function, which is a natural decay pattern in most of physical applications.
When a single bath mode is assumed, i.~e. when the spectral density
function is obtained as a Fourier transform of a single term of Eq.~\eqref{eq:c_t},
the spectral density is given by 
\begin{equation}
\mathcal{C}_{\mathrm{u}}^{\prime\prime}(\omega)=\pi s\omega_{0}^{2}\left[\delta(\omega-\omega_{0})-\delta(\omega+\omega_{0})\right].\label{eq:C_undamped}
\end{equation}
 The reorganization energy in this case is $s\omega_{0}$ and the
lineshape function is 
\begin{equation}
g_{\mathrm{u}}(t)=s\left[\coth\frac{\beta\hbar\omega_{0}}{2}\left(1-\cos\omega_{0}t\right)+\imath\left(\sin\omega_{0}t-\omega_{0}t\right)\right].\label{eq:g_u}
\end{equation}
 This vibrational mode leads to non-decaying, \emph{undamped,} vibrational
motion. Such spectral density given by the $\delta$-functions is
not realistic because dissipation, which accompanies any vibrational motion
due to the molecular interaction with its environment, is neglected.
Damping induces the decay of the correlation function over time and
the corresponding spectral density should have a finite-width smooth
peak. Such \emph{damped} regime is achieved by taking $\gamma<\omega_{0}$
in Eq.~\eqref{eq:C_w_total} and leads to

\begin{equation}
\mathcal{C}_{\mathrm{d}}^{\prime\prime}(\omega)=\frac{2\sqrt{2}\lambda\omega\omega_{0}^{2}\gamma}{\left(\omega^{2}-\omega_{0}^{2}\right)^{2}+2\gamma^{2}\omega^{2}}.\label{eq:C_damped}
\end{equation}
 This spectral density function with reorganization energy $\lambda$
has a peak at $\omega_{0}$ and the peak width is defined by the damping
strength $\gamma$, differently from Eq.~\eqref{eq:C_undamped},
where the peak is the $\delta$-function (Fig.~\ref{fig:Damped_overdamped}a).

The opposite, \emph{overdamped}, regime is usually represented by
the spectral density of a Brownian oscillator, which represents the
fluctuations of the Ornstein-Uhlenbeck process \cite{Breuer-Petruccione}.
This regime can be postulated semi-classically, i.~e. by using the
exponentially decaying classical correlation function $\mathcal{C}_{\mathrm{cl}}(t)=2\lambda k_{\mathrm{B}}T\exp(-\gamma|t|)$.
Its Fourier transform $\mathcal{C}_{\mathrm{cl}}(\omega)=4\gamma\lambda k_{\mathrm{B}}T(\omega^{2}+\gamma^{2})^{-1}$
now represents the even (classical) part $\mathcal{C}^{\prime}(\omega)$
of the total quantum correlation function. Since the classical trajectory
reflects the high temperature limit, we have $\mathcal{C}^{\prime\prime}(\omega)=\frac{1}{2}\beta\hbar\omega\mathcal{C}^{\prime}(\omega)$
and obtain the spectral density representing the classical Brownian
particle \cite{MayBook2011} 
\begin{equation}
\mathcal{C}_{\mathrm{o-sc}}^{\prime\prime}(\omega)=\frac{2\lambda\gamma\omega}{\omega^{2}+\gamma^{2}}\label{eq:o_sc}
\end{equation}
with reorganization energy $\lambda$ and the relaxation rate $\gamma$.
The full quantum correlation function in the frequency domain can
now be constructed by the direct application of the fluctuation-dissipation
relation Eq.~\eqref{eq:fl-dis}. Therefore, we denote bath described
by such spectral density as the \emph{overdamped semi-classical bath}.
At the high-temperature limit the corresponding lineshape function
is obtained as 
\begin{equation}
g_{\mathrm{o-sc}}(t)=\frac{\lambda}{\gamma}\left(\frac{2}{\beta\gamma}-\mathrm{i}\right)(\mathrm{e}^{-\gamma t}+\gamma t-1).\label{eq:gfunc_o_sc}
\end{equation}

Evidently, the overdamped bath can also be obtained from Eq.~\eqref{eq:C_w_total}.
In the limit of $\gamma\gg\omega_{0}$ it yields rather different
expression 
\begin{equation}
\mathcal{C}_{\mathrm{o-q}}^{\prime\prime}(\omega)=\frac{4\lambda\omega\gamma^{3}}{(\omega^{2}+\gamma^{2})^{2}}.\label{eq:o_q}
\end{equation}
 We denote this regime as the \emph{quantum overdamped}. The reorganization
energies are equal to $\lambda$ for both types of the overdamped
spectral density. The semi-classical spectral density function is
equal to $\lambda$ at its maximum at $\omega=\gamma$, while the
quantum function has its maximum value $\mathcal{C}_{\mathrm{o-q}}^{\prime\prime}(\omega_{\mathrm{peak}})=\frac{3\sqrt{3}}{4}\lambda\approx1.3\lambda$,
where $\omega_{\mathrm{peak}}=\sqrt{3}/3\gamma\approx0.58\gamma$
(see Fig.~\ref{fig:Overdamped_q_sc}a). The quantum overdamped spectral
density decays faster at infinity, therefore, it is more suitable
for numerical applications. At the high-temperature limit the corresponding
lineshape function is given by 
\begin{eqnarray}
g_{\mathrm{o-q}}(t) & = & \frac{\lambda}{\gamma}\left(\frac{2}{\beta\gamma}-\mathrm{i}\right)\left(\e^{-\gamma t}\gamma t+2\e^{-\gamma t}-2+\gamma t\right)\label{eq:gfunc_o_q}\\
 &  & +\frac{\lambda\beta}{2}\left(1+\frac{4}{\beta^{2}\gamma^{2}}\right)\left(\mathrm{e}^{-\gamma t}+\gamma t-1\right).\nonumber 
\end{eqnarray}

\section{Spectroscopic observables}

The absorption signals of the two-level system coupled to the bath
are calculated according to Eq.~\eqref{eq:abs-monomer-g} and using
lineshape functions corresponding to different damping regimes (eqs.
\eqref{eq:C_undamped}, \eqref{eq:C_damped}, \eqref{eq:o_sc} or
\eqref{eq:o_q}). The included overdamped modes lead to the Lorentzian
or Gaussian lineshapes of spectral peaks for \emph{fast}, $2\lambda\beta^{-1}\gamma^{-2}\gg1$,
or \emph{slow}, $2\lambda\beta^{-1}\gamma^{-2}\ll1$, decay regimes,
respectively. Fast-decaying modes of molecular vibrations will result
in homogeneous broadening, while the strong coupling to the high-frequency
vibrations will result in vibrational progression in the absorption
spectrum.

In the case when the molecule is coupled to a single high-frequency
mode and a continuum of low-frequency damping modes the spectral density
consists of two parts:

\begin{equation}
\mathcal{C}^{\prime\prime}(\omega)=\mathcal{C}_{\mathrm{o}}^{\prime\prime}(\omega)+\mathcal{C}_{\mathrm{vib}}^{\prime\prime}(\omega).\label{eq:total_sp}
\end{equation}
 Here the (fast) $\mathcal{C}_{\mathrm{o}}^{\prime\prime}(\omega)$
mode corresponds to the \emph{semi-classical} $\mathcal{C}_{\mathrm{o-sc}}^{\prime\prime}(\omega)$
or \emph{quantum} $\mathcal{C}_{\mathrm{o-q}}^{\prime\prime}(\omega)$
overdamped bath. The second term, $\mathcal{C}_{\mathrm{vib}}^{\prime\prime}(\omega)$,
represents spectral density of (slow) molecular vibrations (\emph{undamped}
$\mathcal{C}_{\mathrm{u}}^{\prime\prime}(\omega)$ or \emph{damped}
$\mathcal{C}_{\mathrm{d}}^{\prime\prime}(\omega)$). The total lineshape
function, which is a linear transformation of the spectral density,
then contains two parts as well.

The spectral density and lineshape function formalism presented above
is extended to the photon echo signal calculations for the 2D ES \cite{AbramaviciusMukamelCR2008}.
The response function theory denotes the response function $S(t_{3},t_{2},t_{1})$
as a sum of different pathways of system density operator propagations.
In this study we assume the impulsive limit (laser pulses of infinitesimally
short duration), where the 2D spectrum is defined by the two-dimensional
Fourier transform of the response function over variables $t_{1}$
and $t_{3}$; these then have the meanings of the time delays between
the first and the second laser pulses, and the third laser pulse and
the detection time, respectively \cite{Butkus2010}. $t_{2}$ is the
fixed time delay between the second and the third interaction and
is denoted as the population time. The \emph{rephasing $\vec k_{\mathrm{I}}$
}(Fig.~\ref{fig:Intensities-REPHASING}) and \emph{nonrephasing}
$\vec k_{\mathrm{II}}$ (Fig.~\ref{fig:Intensities-NONREPHASING})
signals are distinguished as the ordering of the interaction with
the first two pulses is changed. The \emph{total} signal is denoted
as the sum of $\vec k_{\mathrm{I}}$ and $\vec k_{\mathrm{II}}$ parts.
Evolution of its diagonal and off-diagonal peaks closely follows the
dynamics of corresponding populations and quantum coherences of the
system density operator. In 2D spectra, fast-decaying modes of molecular
vibrations will result in anti-diagonal peak (homogeneous) broadening,
while the strong coupling to the high-frequency vibrations will result
in multiple oscillating diagonal and off-diagonal peaks reflecting
the configuration of vibrational states \cite{Abramavicius-EPL2007,ButkusValkunas2011}.

For a single two-level system the complete expressions for the rephasing
and nonrephasing contributions of 2D spectrum can be given in terms
of the lineshape functions: 
\begin{eqnarray}
 &  & S_{\vec k_{\mathrm{I}}}(t_{3},t_{2},t_{1})=\frac{1}{2}\e^{-\imath\omega_{0}(t_{3}-t_{1})}\label{eq:mono1}\\
 &  & \quad\times\e^{-g^{*}(t_{1}+t_{2})-g^{*}(t_{1})+g^{*}(t_{1}+t_{2}+t_{3})}\nonumber \\
 &  & \quad\times\e^{\mathrm{Re}\left[g(t_{2}+t_{3})+g(t_{2})-g(t_{3})\right]}\nonumber \\
 &  & \quad\times\cos\left\{ \mathrm{Im}\left[g(t_{2}+t_{3})+g(t_{2})+g(t_{3})\right]\right\} \nonumber 
\end{eqnarray}
 and 
\begin{eqnarray}
 &  & S_{\vec k_{\mathrm{II}}}(t_{3},t_{2},t_{1})=\frac{1}{2}\e^{-\imath\omega_{0}(t_{3}+t_{1})}\label{eq:mono2}\\
 &  & \quad\times\e^{g(t_{1}+t_{2})-g(t_{1})-g(t_{1}+t_{2}+t_{3})}\nonumber \\
 &  & \quad\times\e^{\mathrm{Re}\left[g(t_{2}+t_{3})+g(t_{2})-g(t_{3})\right]}\nonumber \\
 &  & \quad\times\cos\left\{ \mathrm{Im}\left[g(t_{2}+t_{3})+g(t_{2})-g(t_{3})\right]\right\} ,\nonumber 
\end{eqnarray}
 respectively. It should be noted that these expressions are exact
for the harmonic bath and carry no approximation \cite{mukbook}.

For a multi-chromophoric system the response function is a sum of
many contributions, which can be classified to the excited state absorption,
excited state emission and ground state bleaching pathways. Additionally,
the pathways responsible for population transfer are present \cite{Abramavicius-EPL2007}.
Here in eqs. \eqref{eq:mono1} and \eqref{eq:mono2}, there are only two pathways;
the excited state absorption contribution producing negative
peaks in the 2D spectrum is not available due to absence of the third
electronic state.

\subsection{Quantum vs. semi-classical overdamped bath}

\begin{figure}
\includegraphics{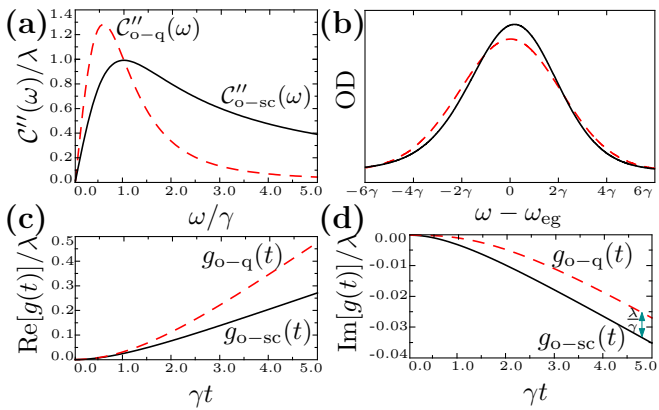}

\caption{\label{fig:Overdamped_q_sc}(a) Overdamped quantum (red dashed line)
and semi-classical (black continuous line) spectral density functions
of an two-level system and corresponding (b) absorption spectra, (c-d)
real and imaginary parts of the lineshape functions.}
\end{figure}

The differences in the two-level system absorption lineshapes obtained
by using solely the \emph{quantum} or \emph{semi-classical} overdamped
bath models ($\mathcal{C}^{\prime\prime}(\omega)\equiv\mathcal{C}_{\mathrm{o}}^{\prime\prime}(\omega)$)
are minor (Fig.~\ref{fig:Overdamped_q_sc}b) despite of a substantial
difference in the shapes of spectral densities (Fig.~\ref{fig:Overdamped_q_sc}a).
The position of the absorption spectra are mainly determined by the
imaginary part of the corresponding lineshape function (Fig.~\ref{fig:Overdamped_q_sc}d):
for the semi-classical bath, it is lower by $\lambda/\gamma$ compared
to the quantum bath at $\gamma t\gg1$. The slopes of the imaginary
parts are, however, both equal. The slopes of the real parts of the lineshape functions,
which determine the absorption linewidth, are different (Fig.~\ref{fig:Overdamped_q_sc}c).
For $g_{\mathrm{o-sc}}(t)$ the function slope at $\gamma t\gg1$
is $\frac{2\lambda}{\beta\gamma}$, while for $g_{\mathrm{o-q}}(t)$
is at least twice larger, $\frac{4\lambda}{\beta\gamma}+\frac{\lambda\beta\gamma}{2}$.
Such differences result in larger lineshape broadening in the case
of the quantum bath.

The shapes of functions describing the spectral peak dynamics of the
rephasing 2D spectra obtained by using the quantum model of the overdamped
bath (Eq.~\eqref{eq:gfunc_o_q}) coincide with the semi-classical
bath simulations (Eq.~\eqref{eq:gfunc_o_sc}) at short population
times, as demonstrated in Fig.~\ref{fig:Intensities-REPHASING}b
and Fig.~\ref{fig:Intensities-REPHASING}d. However, in the nonrephasing
(Fig.~\ref{fig:Intensities-NONREPHASING}b and Fig.~\ref{fig:Intensities-NONREPHASING}d)
spectrum the quantum overdamped bath changes both the amplitudes of
oscillations and the averages of peak intensities. In the 2D spectra
the quantum overdamped bath model produces lineshapes with slightly
smaller broadening. By adjusting the parameters both models can be
tuned to reflect experimental broadenings. The quantum model has a
short tail in the spectral density, which is advantageous for numerical
simulations. 

\subsection{Undamped vs. damped harmonic bath\label{sec:Spectroscopic-signals-in-1}}

The properties of undamped and damped harmonic baths are demonstrated
by assuming the composite spectral density given by Eq.~\eqref{eq:total_sp}
and setting the Huang-Rhys factor of the coherent mode to $s=0.3$.
In absorption, such system demonstrates three well-resolved peaks
of vibrational progression at frequencies $\omega=\omega_{\mathrm{eg}},$
$\omega_{\mathrm{eg}}+\omega_{0}$ and $\omega_{\mathrm{eg}}+2\omega_{0}$
(Fig.~\ref{fig:Damped_overdamped}b).

The case when the composite spectral density is $\mathcal{C}_{\mathrm{o-sc}}^{\prime\prime}(\omega)+\mathcal{C}_{\mathrm{u}}^{\prime\prime}(\omega)$
corresponds to the overdamped semi-classical bath with undamped
vibrations. This is the commonly used approach and is discussed in
detail in the literature \cite{mancal-sperling-jcp2010,egorova-CP2008,Butkus2012}.
The produced vibrational progression in the absorption spectrum is
drawn as the red dashed line in Fig.~\ref{fig:Damped_overdamped}b.
Unsurprisingly, the absorption signal calculated from eqs.~\eqref{eq:abs-monomer-fc}
and \eqref{eq:abs-monomer-g} are then identical.

The effects caused by damping of vibrations are often ignored, however
they may raise significant implications in the nonlinear spectra.
In the linear spectra, comparison of the result obtained by assuming
the spectral density of $\mathcal{C}_{\mathrm{o-sc}}^{\prime\prime}(\omega)+\mathcal{C}_{\mathrm{d}}^{\prime\prime}(\omega)$
(black solid lines in Fig.~\ref{fig:Damped_overdamped}b) to the
undamped vibrations reveals an almost perfect correspondence for $\gamma=0.05\omega_{0}$,
which illustrates a smooth transition from one model to another. It
is also noticeable that the peaks in the progression broaden gradually
when the damping strength $\gamma$ is increased. However, the broadening
is not uniform -- peaks that are at higher energies are broadened
more. This is evident by evaluating peaks width dependence in the
damping strength (Fig.~\ref{fig:Damped_overdamped}c), obtained by
fitting the spectra with multiple Lorentzian functions. As a result,
the total spectral lineshape becomes asymmetric: higher-energy shoulder
of the vibrational progression is reduced due to additional broadening.
Such pattern of broadening is consistent with the lifetime-induced
decay of vibrational coherences at higher excitation energies.

\begin{figure}
\begin{centering}
\includegraphics{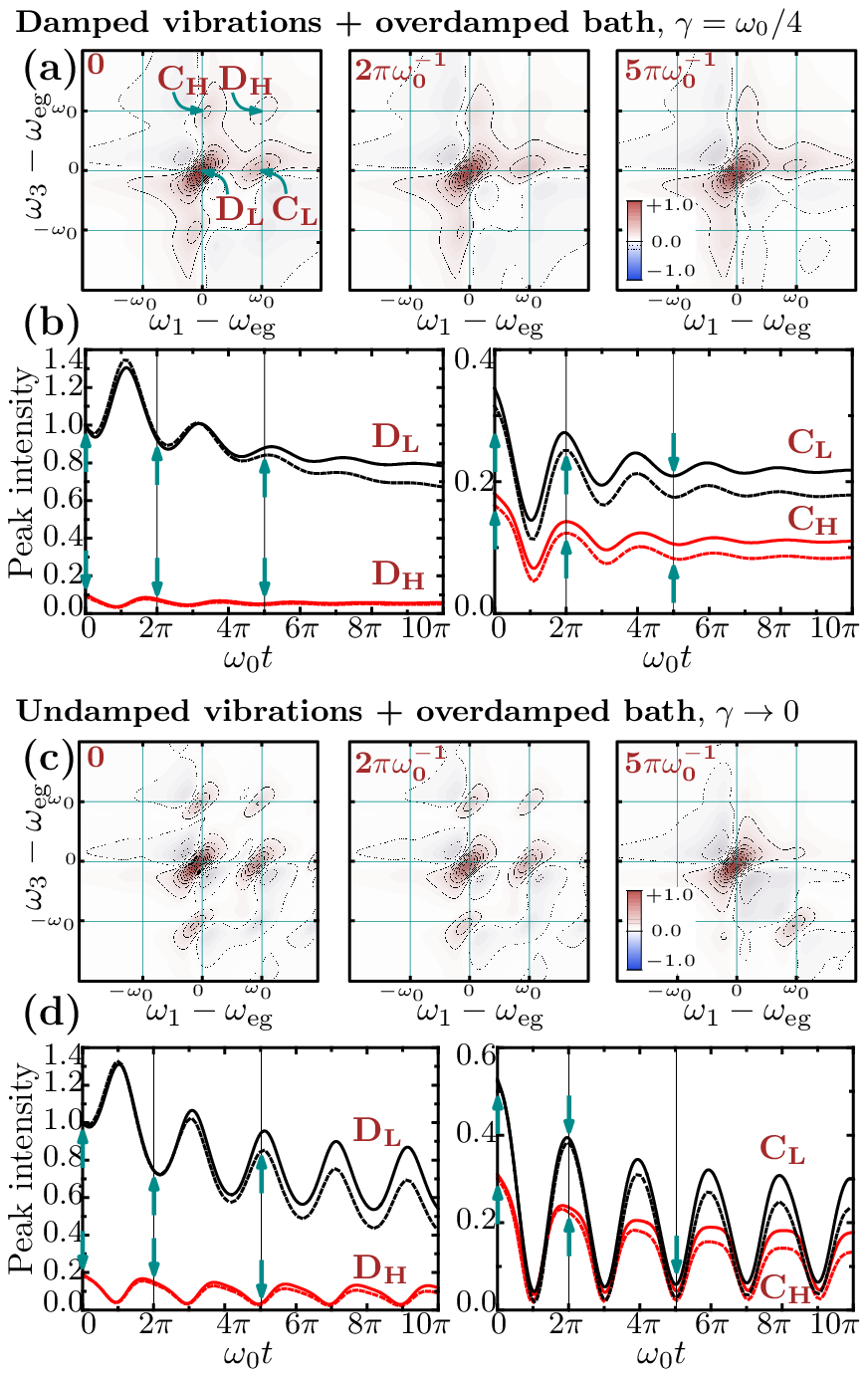} 
\par\end{centering}

\caption{\label{fig:Intensities-REPHASING}2D rephasing spectra at $t_{2}=0$,
$2\pi\omega_{0}^{-1}$ and $5\pi\omega_{0}^{-1}$ with damped (a)
and undamped (c) vibrations on top of the semi-classical bath. Intensities
of the peaks, indicated by `$\mathrm{D_{L}}$' (diagonal lower), `$\mathrm{C_{H}}$'
(cross-peak higher), etc., are depicted as functions of population
time $t_{2}$ in case of damped vibrations (b) and undamped vibrations
(d) with semi-classical (solid lines) and quantum (dashed lines) overdamped
bath. All values are normalized to the maximum of the rephasing spectrum
at $t_{2}=0$.}
\end{figure}

\begin{figure}
\centering{}\includegraphics{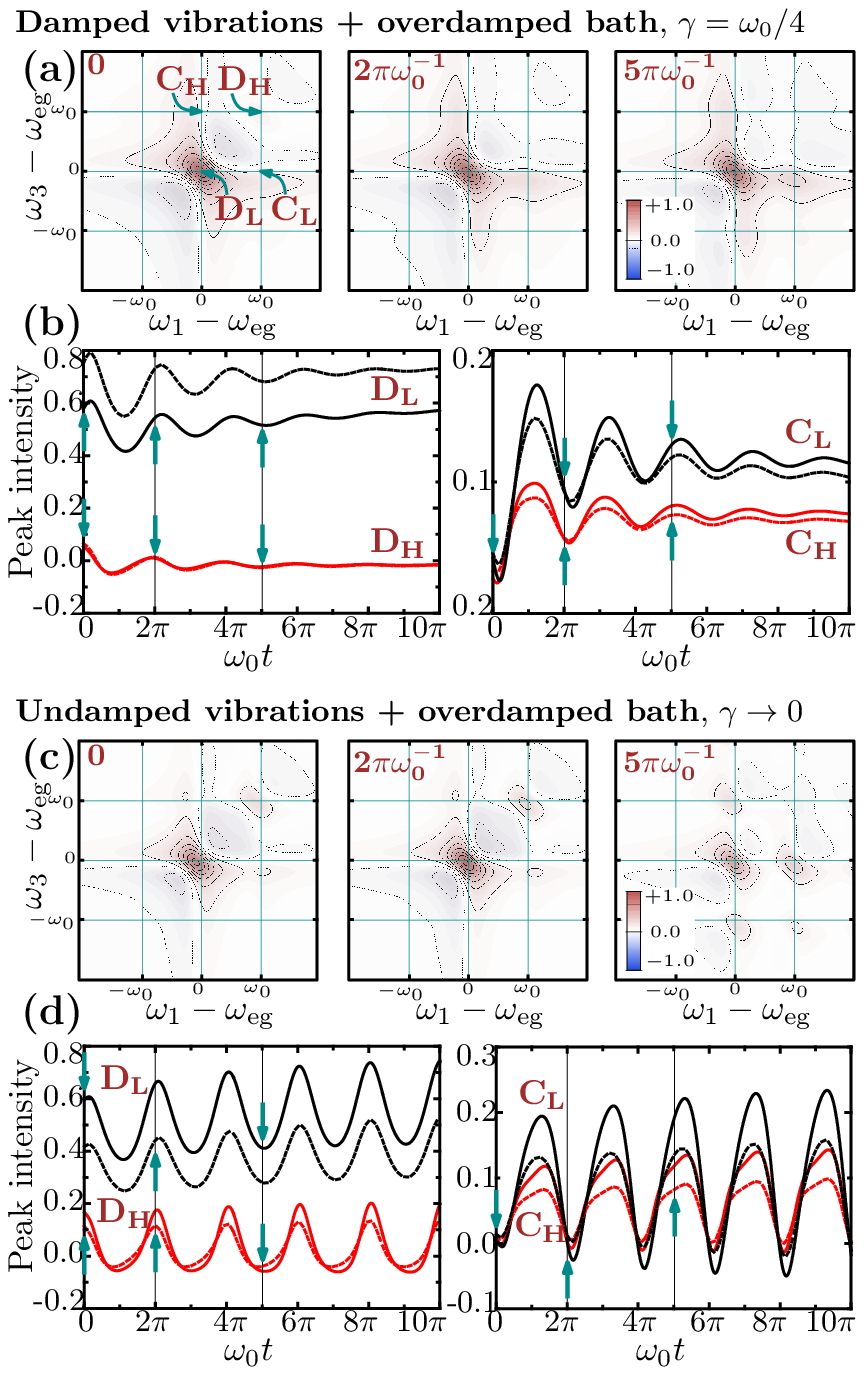}\caption{\label{fig:Intensities-NONREPHASING}2D nonrephasing spectra at $t_{2}=0$,
$2\pi\omega_{0}^{-1}$ and $5\pi\omega_{0}^{-1}$ with damped (a)
and undamped (c) vibrations on top of the semi-classical bath; the
intensities of corresponding peaks in case of damped vibrations (b)
and undamped vibrations (d) with semi-classical (solid lines) and
quantum (dashed lines) overdamped bath. All values are normalized
to the maximum of the corresponding rephasing spectrum at $t_{2}=0$
(in Fig.~\ref{fig:Intensities-REPHASING}).}
\end{figure}

The peak lineshapes in the 2D spectra obtained by using the damped
case show slightly larger broadenings compared to those of the undamped
vibrations both for rephasing and nonrephasing signals (figs.~\ref{fig:Intensities-REPHASING}
and \ref{fig:Intensities-NONREPHASING}). The non-uniform broadening,
as follows from the absorption simulations, shows up in the 2D spectrum,
as well. For the vibrational bath we separately study two cases: the
damped vibrations ($\gamma=\omega_{0}/4$, Fig.~\ref{fig:Intensities-REPHASING}a-b
and Fig.~\ref{fig:Intensities-NONREPHASING}a-b) and the undamped
vibrations ($\gamma\to0$, Fig.~\ref{fig:Intensities-REPHASING}c-d
and Fig.~\ref{fig:Intensities-NONREPHASING}c-d). In both cases the
lineshape function representing the semi-classical overdamped
bath is added to the vibrational part.  The reorganization energy in all cases is the same ($2s\omega_{0}$).

The vibrational wavepacket (coherence) beats show up as temporal oscillations
in the 2D spectrum. Decay of the coherences in case of the damped
vibrations results in decay of cross-peaks (compare spectra for $t_{2}=2\pi\omega_{0}^{-1}$
in Fig\@.~\ref{fig:Intensities-REPHASING}a and Fig.~\ref{fig:Intensities-REPHASING}c):
since all spectra are normalized to the maximum of the rephasing signal
at $t_{2}=0$, the intensities of the cross-peaks and the upper diagonal
peak are evidently lower than those of the main peak at $(\omega_{1},\omega_{3})=(\omega_\mathrm{eg},\omega_\mathrm{eg})$.
The main differences in oscillatory dynamics of peaks in the 2D spectra
are due to the damping-induced decay of coherences in the case of
the damped vibrations.

Considering the phases of oscillations corresponding to different
peaks, it is clearly observed that the model of vibrational bath does
not change the phase relationships of peak oscillations. In general,
the phases of peak oscillations are not the same for all peaks in
rephasing and nonrephasing signals. For example, in the rephasing
signal lower diagonal peak `$\mathrm{D}_\mathrm{L}$' oscillates as the negative
cosine function with the population time $t_{2}$, while all other
peaks oscillate with opposite phase (Fig.~\ref{fig:Intensities-REPHASING}b
and Fig.~\ref{fig:Intensities-REPHASING}d); in the nonrephasing
signal (Fig.~\ref{fig:Intensities-NONREPHASING}b and Fig.~\ref{fig:Intensities-NONREPHASING}d)
both cross-peaks oscillate as the negative cosine, while both diagonal
peaks -- with the opposite phase. Such phase relationships are typical
for oscillations of vibrational coherences and are known as being
strongly dependent on the Huang-Rhys factor; this property ought to
be employed in distinguishing vibrational and electronic coherences
in spectra \cite{Butkus2012}.

\section{Discussion}

We considered the smooth transition from \emph{undamped} through \emph{damped}
to \emph{overdamped }vibrations and their possible manifestation in
the 2D ES signals. For small values of damping parameters ($\gamma<\lambda$)
the undamped vibrational model can be used, but for $\gamma\approx\lambda$,
effects of non-uniform peak broadening in absorption and the damping
strength-dependent decay of coherences in the 2D signal become significant.
We also introduced the quantum overdamped spectral density model which
has a convenient decay character with frequency and can be used to
model broadening effects similar to the semi-classical model of overdamped
Brownian motion.

Understanding of the damping-related effects on the spectra is important
when considering the realistic experimental measurements, where the
mixture of electronic and vibrational quantum beats is expected to
be observed. 
\begin{figure}
\begin{centering}
\includegraphics{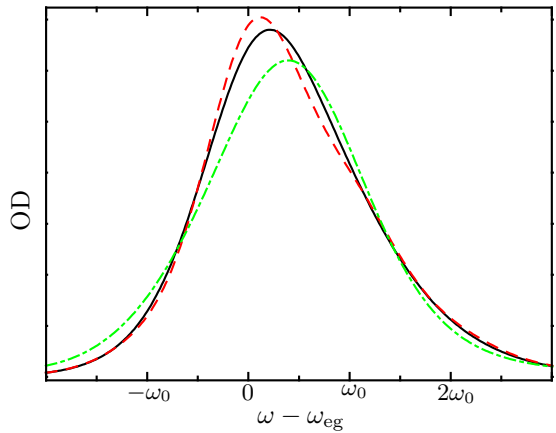} 
\par\end{centering}

\caption{\label{fig:absorptions_3systems}Absorption spectra of the monomer
with damped (solid black line) and undamped (dashed red line) vibrations;
absorption spectrum of a dimer (green dash dotted line).}
\end{figure}

\begin{figure*}
\begin{centering}
\includegraphics{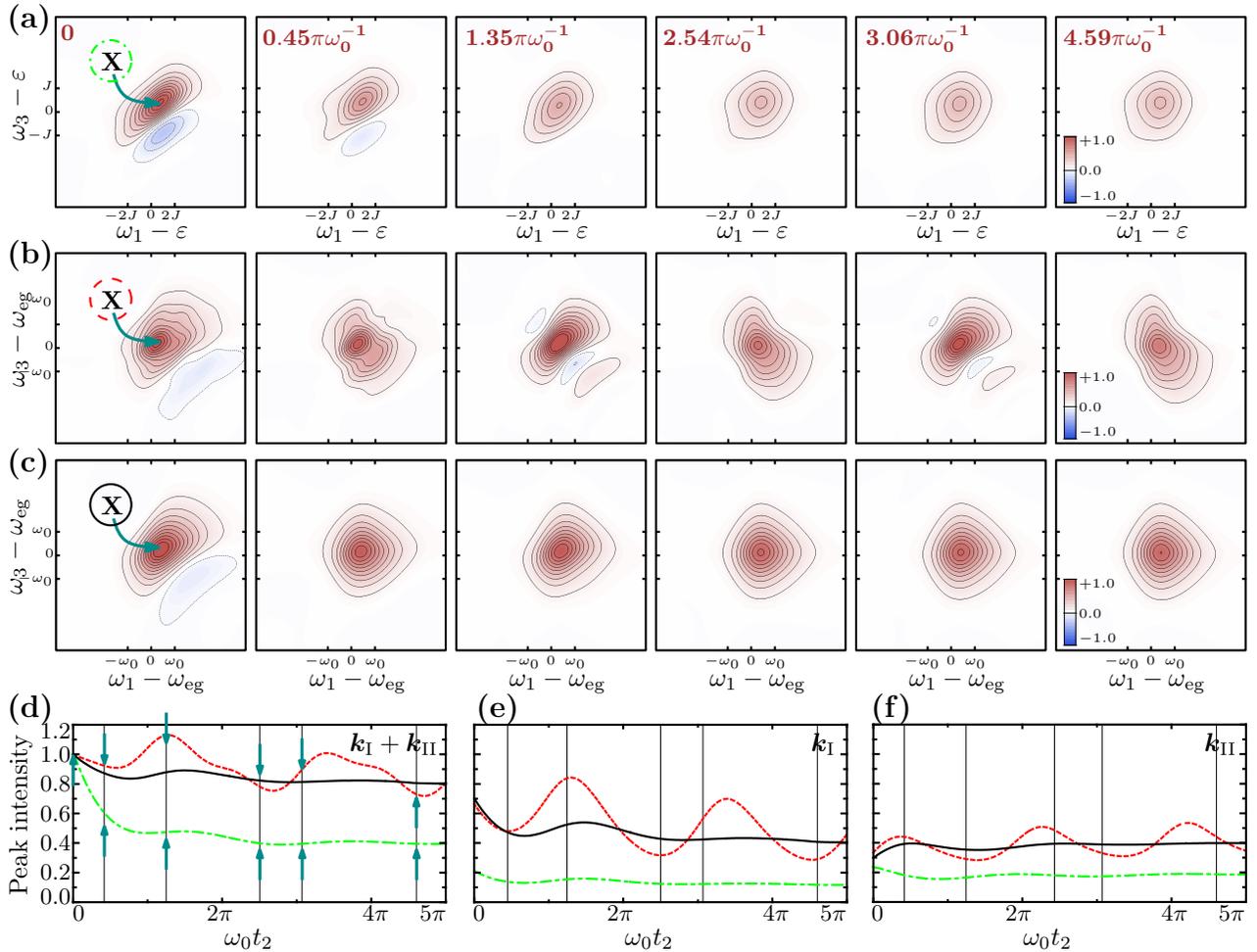} 
\par\end{centering}

\caption{\label{fig:Real-parts-of}Real parts of total ($\vec k_{\mathrm{I}}+\vec{k_{\mathrm{II}}}$)
2D spectra of a dimer (a) and a two-level system with undamped (b)
and damped (c) vibrations at population times $t_{2}=0...4.59\pi\omega_{0}^{-1}$.
Plots of intensities of peaks, indicated by 'X' of total (d), $\vec k_{\mathrm{I}}$
(e) and $\vec k_{\mathrm{II}}$ (f) signals of the monomer with damped
vibrations (solid black lines), undamped vibrations (dashed red lines)
and molecular dimer (dot dashed green lines) are depicted. Spectra
are normalized to the values of intensity maxima at $t_{2}=0$ for
each system separately. }
\end{figure*}

To demonstrate the relations we turn to simulations and comparisons
of model systems that have strong spectral congestion and underlying
system properties may be obscure. To make a constructive parallel
to the typical experimental results, properties of total spectra are
highlighted rather than rephasing and nonrephasing spectra separately.
We compare monomers coupled to (undamped and damped) vibronic resonances
of frequency $\omega_{0}$. The results are compared to the elementary
molecular aggregate -- a molecular homodimer with resonant coupling
$\left|J\right|\equiv\frac{\omega_{0}}{2}$ (Fig.~\ref{fig:absorptions_3systems}).
The parameters of the monomer are the same as in Sec.~\ref{sec:Spectroscopic-signals-in-1}.
The 2D signals of the homodimer are simulated using the conventional response function
theory for molecular excitons as described in Ref.~\cite{Abramavicius2010}.
The site energies of the dimer are taken to be $\varepsilon=\omega_{\mathrm{eg}}+\frac{1}{2}\omega_{0}$
and the inter-dipole angle $\phi=\frac{2}{3}\pi$. Electronic transition
to two excitonic states at $\omega_{\mathrm{eg}}$ and $\omega_{\mathrm{eg}}+\omega_{0}$
are then possible. The higher energy peak is more intensive than the
lower one by factor $\frac{1-\cos\phi}{1+\cos\phi}=3$. For such H-type
dimer \cite{Gelzinis2011} the excited state absorption contribution
will be much stronger in the lower cross-peak, making it negative
in the 2D spectrum (Fig.~\ref{fig:Real-parts-of}a). The inter-dipole
angle also determines the amplitudes of coherence oscillations. Reducing
the angle would increase the amplitude of coherence oscillations,
but decrease the excited state absorption influence on the lower crosspeak.
The spectral density corresponding to the overdamped Brownian oscillator
(overdamped semi-classical bath, Eq.~\eqref{eq:gfunc_o_sc})
is used for the spectral broadening. Its parameters are $\lambda=\frac{3}{5}\omega_{0}$
and $\gamma=\frac{1}{6}\omega_{0}$.

The total spectra of the monomer with undamped vibrations (Fig.~\ref{fig:Real-parts-of}b)
show a typical non-decaying `breathing' behavior of the main peak
\cite{nemeth-sperling-JCP2010,Christensson2011}. It can be characterized
as the anticorrelation of diagonal and antidiagonal widths of the
main peak. Such change of the peak shape is created by the superposition
of $\vec k_{\mathrm{I}}$ and $\vec k_{\mathrm{II}}$ signals, which
have phase-inverted dependence of the peak intensity on $t_{2}$,
as can be seen in Fig.~\ref{fig:Real-parts-of}e and Fig.~\ref{fig:Real-parts-of}f.
The same relation is observed in simulations of systems with well-resolved
peaks as shown in Fig.~\ref{fig:Intensities-REPHASING} and Fig.~\ref{fig:Intensities-NONREPHASING}.

As it was already mentioned, the phase relationships and oscillation
amplitudes of the $\vec k_{\mathrm{I}}$ and $\vec k_{\mathrm{II}}$
signals are Huang-Rhys factor-dependent, therefore, both correlation
and anticorrelation of the diagonal peak-width and intensity is possible
for different systems \cite{Christensson2011}. However, the anticorrelation
of the diagonal width and intensity of any peak in the 2D spectrum
is not available for the electronic dimer \cite{Psliakov-Fleming-JCP2006}.
As follows from our calculations of a two-level system the amplitude
is higher for the $\vec k_{\mathrm{I}}$ signal and, therefore, the
correlation of the diagonal width and intensity of the main peak is
observed both for damped and undamped vibrations. Thus, spectral dynamics
of the vibrational and electronic-only systems are very similar by
this aspect.

The monomer with damped vibrations being analyzed here is an illustration
of the transition from damped to overdamped vibrations. Due to damping,
the main peak experiences only a few `breaths' and the round shape
is established at $t_{2}>2.5\pi\omega_{0}^{-1}$ at least (Fig.~\ref{fig:Real-parts-of}c).
Increasing the damping strength would destroy all vibrational coherences
and it would not be possible to spectrally distinguish it from a general
two-level system coupled to an overdamped bath. However, for a molecular
dimer one can expect it to demonstrate similar spectral characteristics
compared to the monomer with damped vibrations. Indeed, this is true
for the 2D spectra presented in Fig.~\ref{fig:Real-parts-of}a. Due
to the excited state absorption, a negative region below the main
peak appears for short population times. The oscillation pattern of
the peak maximum (green dot-dashed line in Fig.~\ref{fig:Real-parts-of}d-f)
closely resembles the dependence of the monomer with damped vibrations
(black solid line), as well. The decay rates of the signals are very
similar here, but the established values of peak intensities at long
population times are different. This can be explained by the additional
dephasing mechanisms (exciton transfer and lifetime decay) available
for an excitonic system during $t_{2}$.

\section{Conclusions}

Consideration of high-frequency \emph{damped} molecular vibrations
instead of \emph{undamped} is a more realistic description since it
includes dephasing and dissipation that vibronic motion experiences
in solvent. This correction induces non-uniform peak broadening, changes
the position of peaks within the vibrational progression in the absorption
and results in decay of coherences in the 2D spectrum. The significance
of the effect depends on the width of the spectral density function
around the resonance. Therefore, having the coherence decay rate and
the peak broadening quantitatively evaluated from the experiment,
one would be able to estimate the damping parameters of vibrations.

A model of \emph{overdamped} \emph{quantum} bath is suggested in this
paper. It is represented by the spectral density, which is directly
obtained from the quantum-mechanical correlation function of bath
coordinates and not assuming the classical correlation function as
in the \emph{overdamped semi-classical} bath model. The spectral density
of the overdamped quantum bath decays as $\omega^{-3}$ at large frequencies,
which is preferable to $\omega^{-1}$ used in overdamped semi-classical
model. In two-dimensional photon echo signals, it results in increased
homogeneous broadening of peaks. Dynamics of quantum beats is not
affected by this spectral density; however, the ratio of rephasing
and non-rephasing signals changes, as well as peak intensities.

Damping of vibrations causes the decay of coherences; its influence
on the peak shape and correlation/anticorrelation of the diagonal
peak width and intensity is reported to be insignificant compared
to the description of undamped vibrations. Further increase of the
damping strength results in instantaneous disappearance of vibrational
coherences and this limit corresponds to purely overdamped vibrational
motion.

\section*{Acknowledgments}

This research was partially funded by the European Social Fund under
the Global grant measure.

\end{document}